\documentclass[preprint,aps,prd,groupedaddress,showpacs,nofootinbib]{revtex4}

\usepackage{amsmath}
\usepackage{graphicx}
\usepackage{epsfig}
\usepackage{bm}

\def\lQ{\Lambda_{\rm QCD}}
\def\als{\alpha_{\rm s}} 
\def\siml{{\ \lower-1.2pt\vbox{\hbox{\rlap{$<$}\lower6pt\vbox{\hbox{$\sim$}}}}\ }}

\newcommand{\be}{\begin{equation}}
\newcommand{\ee}{\end{equation}}
\newcommand{\bea}{\begin{eqnarray}}
\newcommand{\eea}{\end{eqnarray}}
\newcommand{\nn}{\nonumber}

\begin{document}

\preprint{\tt ANL-HEP-PR-07-2 ~~~ IFUM-886-FT  ~~~ UB-ECM-PF 07/01}

\title{Extraction of $\als$ from radiative $\Upsilon(1S)$ decays}
\author {Nora Brambilla$^1$, Xavier \surname{Garcia i Tormo}$^2$, Joan Soto$^3$ and Antonio Vairo$^1$ \vspace{4mm}}
\affiliation{$^1$ Dipartimento di Fisica dell'Universit\`a di Milano and INFN\\
              via Celoria 16, 20133 Milan, Italy \vspace{4mm}}
\affiliation{$^2$ High Energy Physics Division, Argonne National Laboratory \\  
              9700 South Cass Avenue, Argonne, IL 60439, USA\vspace{4mm}}
\affiliation{$^3$ Dept. d'Estructura i Constituents de la Mat\`eria, Universitat de Barcelona \\ 
              Diagonal 647, E-08028 Barcelona, Catalonia, Spain}

\date{\today}

\begin{abstract}
We improve on a recent determination of $\als$ from 
$\Gamma(\Upsilon(1S) \to \gamma\, X)/\Gamma(\Upsilon(1S) \to X)$
with CLEO data by taking into account color octet contributions and avoiding 
any model dependence in the extraction. We obtain $\als (M_{\Upsilon(1S)})=
0.184^{+0.015}_{-0.014}$,  which corresponds to $\als(M_Z) = 0.119^{+0.006}_{-0.005}$.
\end{abstract}


\maketitle

\section{Introduction}
In the early days of QCD heavy quarkonium states ($H$), bound states of a heavy quark
and a heavy antiquark, provided an ideal probe of the new
theory. Among the interesting features, it looked like the strong coupling
constant $\als$ could be neatly extracted from the ratio $\Gamma (H\rightarrow
\gamma gg)/\Gamma (H\rightarrow ggg)$, for which both the wave function at the
origin and the relativistic corrections cancel out
\cite{Brodsky:1977du,Koller:1978qg}\footnote{See \cite{Brambilla:2004wf} for an update on $\als$ extractions from
heavy quarkonium systems.}.
 The first measurements of $J/\psi$
inclusive radiative decays   by the Mark II collaboration
\cite{Scharre:1980yn} delivered a photon spectrum  not compatible
with the early QCD predictions. To a lesser extent, this was also the case for
bottomonium (see \cite{Wolf:2000pm} and references therein). With the advent
of Non-Relativistic QCD (NRQCD) \cite{Caswell:1985ui,Bodwin:1994jh}, it was
understood that color octet contributions, which were ignored in the early
approaches, become very important in the upper end-point region of the
spectrum \cite{Rothstein:1997ac}. When the color octet contributions are
properly taken into account, a very good description of the photon spectrum can
be obtained from QCD, at least for the $\Upsilon (1S)$ state
\cite{GarciaiTormo:2005ch}. Color octet contributions also affect the ratio
$\Gamma (H\rightarrow \gamma gg)/\Gamma (H\rightarrow ggg)$ and are
parametrically of the same order of the relativistic corrections. They have so
far either been ignored \cite{Besson:2005jv} or estimated to be small
\cite{Hinchliffe:2000yq} in the available extractions of $\als$ from this
ratio. 
In this paper we take into account recent determinations of the $\Upsilon
(1S)$ color octet matrix elements both on the lattice \cite{Bodwin:2005gg} and
in the continuum \cite{GarciaiTormo:2004jw}, which indicate that their
contribution is actually not small. This, together with the good theoretical
description of the photon spectrum \cite{GarciaiTormo:2005ch}, allows for a
consistent extraction of $\als (M_{\Upsilon (1S)} )$ at NLO in the NRQCD
velocity counting. We obtain a precise determination of it from the recent
CLEO data \cite{Besson:2005jv}.

\section{$\als$ extraction}
The experimental value of $R_\gamma\equiv \Gamma(\Upsilon(1S) \to \gamma\,
  X)/\Gamma(\Upsilon(1S) \to X)$, $X$ being hadrons, has been determined most
recently in \cite{Besson:2005jv}\footnote{$\gamma $ stands 
for direct photons only and the contributions $\Upsilon(1S)\to \gamma^\ast \to X$ have 
been subtracted in the denominator.}. We will use only the value obtained 
from the Garcia-Soto (GS) parameterization of data \cite{GarciaiTormo:2005ch}, 
which follows from a QCD calculation. This is
\be
R_\gamma^{\rm exp} = 0.0245 \pm 0.0001 \pm 0.0013,
\label{CLEO}
\ee
where the first error is statistic and the second systematic.

Our starting point is the expression:
\bea
R_\gamma &\equiv& \frac{\Gamma(\Upsilon(1S) \to \gamma\,
  X)}{\Gamma(\Upsilon(1S) \to X)} 
= 
\frac{36}{5}\frac{e_b^2\alpha}{\als} \frac{N}{D},
\label{Rgamma}
\\
N &=& 
1+C_{gg\gamma} \frac{\als}{\pi} 
+ C_{{\cal P}_1(^3S_1)} {\cal R}_{{\cal P}_1(^3S_1)}
+ \frac{\pi}{\als} C_{\gamma\,O_8(^1S_0)} {\cal R}_{O_8(^1S_0)}
\nn\\
& & \qquad\qquad\quad
+ \frac{\pi}{\als} C_{\gamma\,O_8(^3P_0)} {\cal R}_{O_8(^3P_0)}
+ {\cal O}_N(v^3), 
\label{N}
\\
D &=&
1+C_{ggg} \frac{\als}{\pi} 
+ C_{{\cal P}_1(^3S_1)} {\cal R}_{{\cal P}_1(^3S_1)}
+ \frac{\pi}{\als} C_{O_8(^3S_1)} {\cal R}_{O_8(^3S_1)}
+ \frac{\pi}{\als} C_{O_8(^1S_0)} {\cal R}_{O_8(^1S_0)}
\nn\\
& & \qquad\qquad\quad
+ \frac{\pi}{\als} C_{O_8(^3P_0)} {\cal R}_{O_8(^3P_0)}
+ {\cal O}_D(v^3), 
\label{D}
\eea
where $n_f=4$ is the number of active flavors, 
$\alpha$ the fine structure constant, $e_b=-1/3$  the bottom quark
electromagnetic charge, $\als = \als(M_{\Upsilon(1S)})$ is the strong coupling
constant calculated at the $\Upsilon(1S)$ mass, $M_{\Upsilon(1S)} = 9.46$ GeV, 
$C_{{\cal P}_1(^3S_1)} = - (19\pi^2 -132)/(12\pi^2
-108)$ \cite{Keung:1982jb},  $C_{\gamma\,O_8(^1S_0)} = 27/(4\pi^2 -36)$ \cite{Maltoni:1998nh}, 
$C_{\gamma\,O_8(^3P_0)} = 189/(4\pi^2 -36)$ \cite{Maltoni:1998nh}, 
$C_{O_8(^3S_1)} = 81 n_f /( 20\pi^2 -180)$ \cite{Petrelli:1997ge}, $C_{O_8(^1S_0)} =  81/(8\pi^2
-72)$ \cite{Petrelli:1997ge}, $C_{O_8(^3P_0)} = 567/(8\pi^2 -72)$ \cite{Petrelli:1997ge}, 
$C_{gg\gamma} = - 1.71$ (for $n_f=4$)
\cite{Kramer:1999bf,Mackenzie:1981sf}, $C_{ggg} = 3.79 \pm 0.54$ (for $n_f=4$)
\cite{Bodwin:1994jh,Mackenzie:1981sf},  ${\cal R}_{O} = \langle \Upsilon(1S)| O | \Upsilon(1S)\rangle/
(m_b^{\Delta d} \, \langle \Upsilon(1S)| O_1(^3S_1)| \Upsilon(1S)\rangle)$, where 
$\Delta d$ is the difference in dimension between the operators $O$ and
$O_1(^3S_1)$, $m_b$ is the bottom mass,  and the $\langle \Upsilon(1S)| O |
\Upsilon(1S)\rangle$ are NRQCD decay matrix
elements \cite{Bodwin:1994jh}. If we adopt the counting of
\cite{Bodwin:1994jh} and $\als/\pi \sim v^2$, then the expansions (\ref{N}) and (\ref{D}) are
valid up to order $v^2$. ${\cal O}_N(v^3)$ and ${\cal O}_D(v^3)$ account for
higher-order corrections of order $v^3$. In the following we will assume $v^2 = 0.08$.

In order to obtain a sensible extraction, the ratios of NRQCD matrix elements ${\cal R}$
must be correctly estimated. ${\cal R}_{{\cal P}_1(^3S_1)}$ can be related to
the binding energy \cite{Gremm:1997dq,Brambilla:2002nu}. ${\cal R}_{O_8(^1S_0)}$ and ${\cal
R}_{O_8(^3S_1)}$ have been calculated on the lattice
\cite{Bodwin:2005gg}. ${\cal R}_{O_8(^1S_0)}$ and ${\cal R}_{O_8(^3P_0)}$ have
been estimated in the continuum \cite{GarciaiTormo:2004jw}. The continuum
calculation and one of the lattice calculations of ${\cal R}_{O_8(^1S_0)}$ are
compatible. We will present two different extractions: C (for continuum) and L
(for lattice). Extraction C uses all the weak-coupling expressions available,
in the same way they were used for the description of the photon spectrum
\cite{GarciaiTormo:2005ch}, and the lattice calculation of
\cite{Bodwin:2005gg} for ${\cal R}_{O_8(^3S_1)}$. Extraction L uses all the
lattice calculations available, and NRQCD velocity scaling to estimate ${\cal
R}_{O_8(^3P_0)}$.  In both extractions, we do not expand the ${\cal
O} (v^2)$ terms in $D$: it turns out that even though they are individually
small they add up to give a contribution comparable to one; we will comment 
on this in Sec. \ref{secdiscussion}.

\subsection{Extraction L (for lattice)}
Concerning the ratios ${\cal R}_{O}$, we will take them in the following ranges:
\bea
 0 \le &{\cal R}_{O_8(^1S_0)}& \le 4.8 \times 10^{-3}, 
\label{R81S0}
\\
 0 \le &{\cal R}_{O_8(^3S_1)}& \le 1.6 \times 10^{-4}, 
\label{R83S1}
\\
 - 2.4 \times 10^{-4}  \le &{\cal R}_{O_8(^3P_0)}& \le 2.4 \times 10^{-4}, 
\label{R81P1}
\\
 - 0.16  \le &{\cal R}_{{\cal P}_1(^3S_1)}& \le 0.
\label{R13S1}
\eea
Equations (\ref{R81S0}) and (\ref{R83S1}) correspond to the maximum values
obtained in the lattice calculations \cite{Bodwin:2005gg} taken with a 100\% uncertainty.
Equation (\ref{R81P1}) follows from the naive counting 
\be
|{\cal R}_{O_8(^3P_0)}| =
\frac{1}{9} \left| \frac{\langle \eta_b| O_8(^1P_1) | \eta_b\rangle}{
m_b^2 \, \langle \Upsilon(1S)| O_1(^3S_1)| \Upsilon(1S)\rangle} \right|
\approx 
\frac{1}{9} \times \frac{v^4}{2N_c},
\label{scalingspin}
\ee
taken with a 100\%  uncertainty. 
The first equality is due to spin symmetry
and is valid at leading order in the velocity expansion, in the second one 
we have evidenced the color factor $1/(2N_c)$ \cite{Bodwin:2005gg} 
($N_c = 3$ is the number of colors). 

Equation (\ref{R13S1}) follows from the Gremm--Kapustin relation
\cite{Gremm:1997dq,Brambilla:2002nu} in the weak-coupling regime,
\be
{\cal R}_{{\cal P}_1(^3S_1)} = \frac{E_{\rm bin}}{m_b} \approx -v^2,
\label{GK}
\ee
taken with a 100\% uncertainty. $E_{\rm bin}$ stands for the binding energy.
The operators ${\cal O}_N(v^3)$ and ${\cal  O}_D(v^3)$ are taken in the range 
\be
- 0.04 \le {\cal O}_N(v^3),{\cal  O}_D(v^3) \le 0.04 \,,
\label{OND}
\ee
which encompasses both ${\cal O}(v^3)$ and ${\cal O}(\als^2)$ corrections.

The fine structure constant is taken at the $\Upsilon(1S)$ mass
\be
\alpha(M_{\Upsilon(1S)}) = \frac{1}{132}.
\label{alphaem}
\ee

We evaluate the theoretical side of Eq. (\ref{Rgamma}) as it stands, without further expansions.
Taking the central values for $C_{ggg}$ and in Eqs. (\ref{R81S0})-(\ref{R13S1}), (\ref{OND}) and (\ref{CLEO}), 
we obtain
\be
\als(M_{\Upsilon(1S)}) = 0.1885 \;.
\ee
The uncertainty on $\als$ induced by a given parameter is evaluated 
by varying it in the range and keeping all other parameters at their central values.
We obtain
\bea
&& \delta_{C_{ggg}} \als = 0.0025,
\\
&& \delta_{{\cal R}_{O_8(^1S_0)}} \als  = 0.0047,
\\
&& \delta_{{\cal R}_{O_8(^3S_1)}} \als = 0.0019,
\\
&& \delta_{{\cal R}_{O_8(^3P_0)}} \als = 0.0032,
\\
&& \delta_{{\cal R}_{{\cal P}_1(^3S_1)}} \als = 0.0106,
\\
&& \delta_{{\cal R}_{{\cal O}_N(v^3)}} \als  = 0.0041,
\\
&& \delta_{{\cal R}_{{\cal O}_D(v^3)}} \als = 0.0031,
\\
&& \delta_{R_\gamma^{\rm exp}} \als = 0.0089.
\eea
We sum up linearly the errors $\delta_{{\cal R}_{O_8(^1S_0)}}$ and $
\delta_{{\cal R}_{O_8(^3S_1)}}$, which are correlated, and then all the
errors quadratically, obtaining
\be
\als(M_{\Upsilon(1S)}) = 0.189 \pm 0.017 \,.
\label{firstalphas}
\ee

The dominant error comes from the uncertainty in ${\cal R}_{{\cal P}_1(^3S_1)}$.
We can reduce this uncertainty, by noticing that for  ${\cal R}_{{\cal
    P}_1(^3S_1)}$ we have an explicit expression, Eq. (\ref{GK}), that 
we have only partially exploited. Indeed, in the weak-coupling regime,
the exact form of $E_{\rm bin}$ is known. At the order we are interested
in, it holds that 
\be\label{ebinwk}
\frac{E_{\rm bin}}{m_b} = - \frac{(C_F\als)^2}{4}, \quad C_F = \frac{4}{3},
\ee
where $\als$ is evaluated at the scale $M_{\Upsilon(1S)} C_F \als/2$, the typical
momentum-transfer scale in a Coulombic bound state. From Eq. (\ref{firstalphas}), 
we obtain:
\be
\als(M_{\Upsilon(1S)} C_F \als/2) = 0.311 \pm 0.032\,,
\ee
which gives 
\be 
{\cal R}_{{\cal P}_1(^3S_1)} = \frac{E_{\rm bin}}{m_b} = - (0.043^{+0.009}_{-0.008}).
\ee
Using this value for ${\cal R}_{{\cal P}_1(^3S_1)}$ and performing again the
above calculation we obtain the new central value 
\be
\als(M_{\Upsilon(1S)}) = 0.183 \,,
\ee
and the new uncertainties 
\bea
&& \delta_{C_{ggg}} \als = 0.0026,
\\
&& \delta_{{\cal R}_{O_8(^1S_0)}} \als  = 0.0040,
\\
&& \delta_{{\cal R}_{O_8(^3S_1)}} \als = 0.0026,
\\
&& \delta_{{\cal R}_{O_8(^3P_0)}} \als = 0.0027,
\\
&& \delta_{{\cal R}_{{\cal P}_1(^3S_1)}} \als = 0.0014,
\\
&& \delta_{{\cal R}_{{\cal O}_N(v^3)}} \als  = 0.0044,
\\
&& \delta_{{\cal R}_{{\cal O}_D(v^3)}} \als = 0.0033,
\\
&& \delta_{R_\gamma^{\rm exp}} \als = 0.0085.
\eea
Summing up the errors like before, we obtain as our best estimate
\be
\als(M_{\Upsilon(1S)}) = 0.183 \pm 0.013 \,.
\label{finalalphas}
\ee
This corresponds to a strong coupling constant at the $M_Z$ mass of 
\be
\als(M_Z) = 0.119 \pm 0.005 \,.
\label{alphasMZ}
\ee

\subsection{Extraction C (for continuum)}
In a weak coupling analysis, $\langle \Upsilon(1S)| O_1(^3S_1)|
\Upsilon(1S)\rangle$ can be calculated in perturbation theory of $\als
(m_bv)$. A NNLO expression is necessary at ${\cal O} (v^2)$
\cite{Melnikov:1998ug,Penin:1998kx}\footnote{We count $\als$ at the soft scale as order $v$.}. 
In order to follow the same procedure as
in \cite{GarciaiTormo:2005ch}, we multiply the leading order term in the decay
widths by the NNLO expression for $\langle \Upsilon(1S)| O_1(^3S_1)|
\Upsilon(1S)\rangle$ and the $\als$ correction to the decay widths by the LO
expression for that matrix element. If we factor out the NNLO matrix element,
this produces a shift $N\rightarrow N + \delta N$ and $D\rightarrow D + \delta
D$ in (\ref{N}) and (\ref{D}):
\bea 
\delta N  & = & C_{gg\gamma}\frac{\als}{\pi}\delta \,,
\\ 
\delta D  & = & C_{ggg}\frac{\als}{\pi}\delta\,, 
\eea 
with 
\be 
\delta=\frac{\langle \Upsilon(1S)|
O_1(^3S_1)| \Upsilon(1S)\rangle_{\rm LO}}{\langle \Upsilon(1S)| O_1(^3S_1)|
\Upsilon(1S)\rangle_{\rm NNLO}}-1 \,.
\ee 
For the central values of the objects below
we take exactly the same ones used in \cite{GarciaiTormo:2005ch}, namely 
\bea
\delta & = & -0.57\, ,\\ 
{\cal R}_{{\cal P}_1(^3S_1)} & = & -0.015\, ,\\ 
{\cal R}_{O_8(^1S_0)} & = & 0.0012\, ,\\ 
{\cal R}_{O_8(^3P_0)} & = & 0.0011\, . 
\label{RcontP}
\eea
For ${\cal R}_{O_8(^3S_1)}$, we use the hybrid algorithm
output of the lattice calculation \cite{Bodwin:2005gg}\footnote{The hybrid
algorithm is selected because compares well with the continuous estimate for
${\cal R}_{O_8(^1S_0)}$.},  
\be 
{\cal R}_{O_8(^3S_1)}=8\times 10^{-5}\, .  
\ee
Using those values, we obtain
\begin{equation}
\als(M_{\Upsilon(1S)}) =  0.185\;.
\end{equation}

In order to associate errors to these central values, we move the values 
of the objects below in the following ranges:
\bea
0.18 \le & \als (m_bv) &\le 0.38\,,
\\
0.32 \le & \als (m_bv^2) &\le 1.3 \,,
\\
0 \le &{\cal R}_{O_8(^3S_1)}& \le 1.6 \times 10^{-4}\,.
\eea
The wide variation range of $\als (m_bv)$ and $\als (m_bv^2)$ 
is expected to account for ${\cal O} (\lQ )$ uncertainties in 
the weak coupling estimates of $O_8(^1S_0)$ and $O_8(^3P_0)$. The upper limit
of $O_8(^3S_1)$ corresponds to twice the largest value obtained 
using the lattice algorithms in \cite{Bodwin:2005gg}. 

The uncertainty on $\als$ induced by a given parameter is evaluated 
by varying it in the range and keeping all other parameters at their central values.
We obtain
\bea
&& \delta_{C_{ggg}} \als = 0.0009,
\\
&& \delta_{\als (m_bv)} \als = ^{+0.0006}_{-0.0064},
\\
&& \delta_{\als (m_bv^2)} \als  = ^{+0.0083}_{-0.0076},
\\
&& \delta_{{\cal R}_{O_8(^3S_1)}} \als = 0.0016,
\\
&& \delta_{{\cal R}_{{\cal O}_N(v^3)}} \als  = ^{+0.0035}_{-0.0034}\, ,
\\
&& \delta_{{\cal R}_{{\cal O}_D(v^3)}} \als = ^{+0.0026}_{-0.0025}\, ,
\\
&& \delta_{R_\gamma^{\rm exp}} \als = 0.01.
\eea
We assume these errors to be independent and sum them up quadratically, obtaining
\be
\als(M_{\Upsilon(1S)}) =  0.185^{+0.014}_{-0.015}  \,.
\ee
This corresponds to a strong coupling constant at the $M_Z$ mass of 
\be
\als(M_Z) = 0.120^{+0.005}_{-0.006} \,.
\ee

\section{Discussion}
\label{secdiscussion}
We have presented two extractions of $\als$ at NLO in the NRQCD velocity
counting, the main differences being the values assigned to the
NRQCD matrix elements. The two outcomes are very close so we take the average
as our final central value. Since the two extractions are not completely
independent we take as our error the range of the two determinations.
Then our final value is  
\be 
\als (M_{\Upsilon(1S)})= 0.184^{+0.015}_{-0.014}\,,
\ee
which corresponds to 
\be 
\als (M_Z) = 0.119^{+0.006}_{-0.005}\,,
\ee 
which is very close to the central value of the PDG \cite{Yao:2006px} with competitive errors. The key
ingredients to get these numbers are the precise CLEO data
\cite{Besson:2005jv}, the use of a QCD calculation (called GS model in
\cite{Besson:2005jv}) to extrapolate the photon spectrum at low $z$, and
accurate estimates of the color octet matrix elements, which have been possible
thanks to recent lattice and continuum estimates. Concerning the matrix elements, our
results are rather insensitive to the values of $O_8(^1S_0)$ and $O_8(^3P_0)$
(note that the upper limit given in Eq. (\ref{R81P1}) for ${\cal
  R}_{O_8(^3P_0)}$, based on the scaling (\ref{scalingspin}), is 
smaller by a factor five than the continuum estimate (\ref{RcontP})),
but would be sensitive to large values of $O_8(^3S_1)$. However, the lattice values for 
$O_8(^3S_1)$, which we have used, turn out to be much smaller than what NRQCD velocity
scaling rules suggest, and do not have a major impact in our results. 

How reliable is our extraction? Our determination is valid at next-to-leading order 
in $\als(m_b)$ and in $v^2$. At this order, terms corresponding to new
qualitative features appear (radiative, relativistic, octet
corrections), each of them of natural size, but whose sum is of order one and hence large.
This is not unusual. It is crucial, however, that higher-order corrections, those that
we have generically labeled as ${\cal O}_N$ and ${\cal O}_D$, are small. 
This is expected because higher-order corrections do not
introduce new qualitative features.
In  Ref. \cite{Bodwin:2002hg}, higher-order corrections in the velocity expansion 
can be found for ${\cal O}_D$. These are not the complete set of corrections 
entering in  ${\cal O}_D$, since higher-order $\als$ corrections to the lowest 
operators are missing. Anyway, using the analogous of Eq. (\ref{GK}) to estimate the matrix elements 
and taking $E_{\rm bin}/m_b \approx - 0.04$, the corrections calculated in \cite{Bodwin:2002hg}
amount to about 0.02, which is consistent with the estimate (\ref{OND}).
Analogous corrections for ${\cal O}_N$ are not known. At present, the main uncertainty 
in our extraction  of $\als$ comes from  the systematic uncertainties in ${R_\gamma^{\rm exp}}$.

Let us next compare our extraction to two previous related ones
\cite{Besson:2005jv,Hinchliffe:2000yq}.

Concerning the extraction of the CLEO paper \cite{Besson:2005jv}, there are two main
differences. (i) On the theoretical side an old formula was used there
\cite{Mackenzie:1981sf}, in which the NRQCD color octet operators were
ignored. This introduces large theoretical uncertainties. In practice,
however, we find that numerically they are not so important for the final
result. (ii) For the total radiative width, two numbers are quoted depending on
whether the so called Field model \cite{Field:1983cy} or the GS model, which
is in fact a QCD calculation, are used for the extrapolation of the photon
spectrum at low $z$. The final number is given as the average of the two
procedures. We believe that the use of the Field model, which uses a parton shower
Monte Carlo technique to incorporate the effects of gluon radiation by the
outgoing gluons in the decay, introduces an unnecessary model dependence that 
moves the actual central value and artificially increases the errors. Our
final results are similar to the ones presented in \cite{Besson:2005jv} for
the GS model.

Concerning the extraction of \cite{Hinchliffe:2000yq}, which is used by 
the PDG \cite{Yao:2006px}, there are three main differences. 
(i) On the theoretical side, the color octet NRQCD matrix elements are ignored in 
$\Gamma(\Upsilon(1S)\to\gamma X)$, whereas  we find that they contribute 
between $30\%$ and $100\%$.\footnote{The color octet contributions in $\Gamma(\Upsilon(1S)\to X)$ 
are estimated to be $9\%$, whereas ours turns out to be between $50\%$ and $160\%$.}
(ii) Older data are used, which are fully consistent with, but not as
precise as, the more recent ones, and an older analysis \cite{Nemati:1996xy},
which relies on the Field model for extrapolations to low $z$. 
(iii) The extraction is actually done from $\Gamma(\Upsilon(1S)\to\gamma X)/\Gamma (\Upsilon(1S)\to l^+ l^-)$.
We believe, contrary to a statement in
\cite{Hinchliffe:2000yq}, that the latter increases rather than decreases the
theoretical uncertainties associated to color octet operators. Indeed, whereas
the ratio radiative/total has the same color octet operators in the numerator
and denominator except for one, the ratio radiative/leptonic (total/leptonic)
has two (three) different color octet operators in the numerator and
denominator. Furthermore, the leptonic width is known to suffer from large higher order
corrections in $\als$ (see \cite{Pineda:2006ri} for a recent discussion), which introduces further uncertainties.

\section{Conclusions}
We have improved on current determinations of $\als (M_{\Upsilon(1S)})$ from
radiative decays of $\Upsilon(1S)$ by avoiding any model dependence and by
taking into account recent estimates of color octet operators. The value we
obtain, $\als (M_Z) = 0.119^{+0.006}_{-0.005}$, is close to the PDG average with competitive errors.

\begin{acknowledgments}
X.G.T., J.S. and A.V. thank Dave Besson for discussions.
A.V. thanks Geoffrey Bodwin for communications.
N. B. thanks Matthias Jamin for discussions.
Part of this work has been carried out at ECT*, Trento, in August 2006, during the program ``Heavy
quarkonium and related heavy quark systems''. 
We acknowledge financial support from the cooperation agreement 
MEC-INFN. N.B., J.S. and A.V. are also supported by the network Flavianet (EU)
MRTN-CT-2006-035482 and J.S. by MEC (Spain) grant CYT FPA
2004-04582-C02-01 and the CIRIT (Catalonia) grant 2005SGR00564. 
The work of X.G.T. was also supported in part by the U.S. Department of
Energy, Division of High Energy Physics, under contract DE-AC02-06CH11357. 
A.V. acknowledges the financial support obtained inside the Italian
MIUR program  ``incentivazione alla mobilit\`a di studiosi stranieri e
italiani residenti all'estero''. 
\end{acknowledgments}


\begin{thebibliography}{99}

\bibitem{Brodsky:1977du}
 S.~J.~Brodsky, D.~G.~Coyne, T.~A.~DeGrand and R.~R.~Horgan,
 Phys.\ Lett.\ B {\bf 73}, 203 (1978).

\bibitem{Koller:1978qg}
 K.~Koller and T.~Walsh,
 Nucl.\ Phys.\ B {\bf 140}, 449 (1978).


\bibitem{Brambilla:2004wf}
  N.~Brambilla {\it et al.},
  Heavy quarkonium physics,
  CERN-2005-005, (CERN, Geneva, 2005)
  [arXiv:hep-ph/0412158].


\bibitem{Scharre:1980yn}
 D.~L.~Scharre {\it et al.},
 Phys.\ Rev.\ D {\bf 23}, 43 (1981).

\bibitem{Wolf:2000pm}
 S.~Wolf,
 Phys.\ Rev.\ D {\bf 63}, 074020 (2001) 
 [arXiv:hep-ph/0010217].

\bibitem{Caswell:1985ui}
 W.~E.~Caswell and G.~P.~Lepage,
 Phys.\ Lett.\ B {\bf 167}, 437 (1986).

\bibitem{Bodwin:1994jh}
 G.~T.~Bodwin, E.~Braaten and G.~P.~Lepage,
 Phys.\ Rev.\ D {\bf 51}, 1125 (1995) 
 [Erratum-ibid.\ D {\bf 55}, 5853 (1997)]
 [arXiv:hep-ph/9407339].

\bibitem{Rothstein:1997ac}
 I.~Z.~Rothstein and M.~B.~Wise,
 Phys.\ Lett.\ B {\bf 402}, 346 (1997) 
 [arXiv:hep-ph/9701404].

\bibitem{GarciaiTormo:2005ch}
 X.~Garcia i Tormo and J.~Soto,
 Phys.\ Rev.\ D {\bf 72}, 054014 (2005) 
 [arXiv:hep-ph/0507107].

\bibitem{Besson:2005jv}
 D.~Besson {\it et al.}  [CLEO Collaboration],
 Phys.\ Rev.\ D {\bf 74}, 012003 (2006) 
 [arXiv:hep-ex/0512061].

\bibitem{Hinchliffe:2000yq}
 I.~Hinchliffe and A.~V.~Manohar,
 Ann.\ Rev.\ Nucl.\ Part.\ Sci.\  {\bf 50}, 643 (2000) 
 [arXiv:hep-ph/0004186].


\bibitem{Bodwin:2005gg}
 G.~T.~Bodwin, J.~Lee and D.~K.~Sinclair,
 Phys.\ Rev.\ D {\bf 72}, 014009 (2005) 
 [arXiv:hep-lat/0503032].

\bibitem{GarciaiTormo:2004jw}
 X.~Garcia i Tormo and J.~Soto,
 Phys.\ Rev.\ D {\bf 69}, 114006 (2004) 
 [arXiv:hep-ph/0401233].

\bibitem{Keung:1982jb}
  W.~Y.~M.~Keung and I.~J.~Muzinich,
  Phys.\ Rev.\ D {\bf 27}, 1518 (1983). 

\bibitem{Maltoni:1998nh}
 F.~Maltoni and A.~Petrelli,
 Phys.\ Rev.\ D {\bf 59}, 074006 (1999) 
 [arXiv:hep-ph/9806455].

\bibitem{Petrelli:1997ge}
  A.~Petrelli, M.~Cacciari, M.~Greco, F.~Maltoni and M.~L.~Mangano,
  Nucl.\ Phys.\ B {\bf 514}, 245 (1998) 
  [arXiv:hep-ph/9707223].

\bibitem{Kramer:1999bf}
 M.~Kr\"amer,
 Phys.\ Rev.\ D {\bf 60}, 111503 (1999) 
 [arXiv:hep-ph/9904416].

\bibitem{Mackenzie:1981sf}
 P.~B.~Mackenzie and G.~P.~Lepage,
 Phys.\ Rev.\ Lett.\  {\bf 47}, 1244 (1981) 

\bibitem{Gremm:1997dq}
 M.~Gremm and A.~Kapustin,
 Phys.\ Lett.\ B {\bf 407}, 323 (1997) 
 [arXiv:hep-ph/9701353].

\bibitem{Brambilla:2002nu}
  N.~Brambilla, D.~Eiras, A.~Pineda, J.~Soto and A.~Vairo,
  Phys.\ Rev.\ D {\bf 67}, 034018 (2003)
  [arXiv:hep-ph/0208019].

\bibitem{Melnikov:1998ug}
 K.~Melnikov and A.~Yelkhovsky,
 Phys.\ Rev.\ D {\bf 59}, 114009 (1999) 
 [arXiv:hep-ph/9805270].

\bibitem{Penin:1998kx}
 A.~A.~Penin and A.~A.~Pivovarov,
 Nucl.\ Phys.\ B {\bf 549}, 217 (1999) 
 [arXiv:hep-ph/9807421].

\bibitem{Yao:2006px}
 W.~M.~Yao {\it et al.}  [Particle Data Group],
 J.\ Phys.\ G {\bf 33}, 1 (2006). 

\bibitem{Bodwin:2002hg}
  G.~T.~Bodwin and A.~Petrelli ,
  Phys.\ Rev.\ D {\bf 66}, 094011 (2002)
  [arXiv:hep-ph/0205210].

\bibitem{Field:1983cy}
 R.~D.~Field,
 Phys.\ Lett.\ B {\bf 133}, 248 (1983). 

\bibitem{Nemati:1996xy}
 B.~Nemati {\it et al.}  [CLEO Collaboration],
 Phys.\ Rev.\ D {\bf 55}, 5273 (1997) 
 [arXiv:hep-ex/9611020].

\bibitem{Pineda:2006ri}
  A.~Pineda and A.~Signer,
  Nucl.\ Phys.\ B {\bf 762}, 67 (2007) 
  [arXiv:hep-ph/0607239].

\end{thebibliography}
\end{document}